\documentclass{sig-alternate-10pt}

\usepackage{url}
\usepackage{graphicx}
\usepackage{wrapfig}
\usepackage{multirow}
\usepackage{textcomp}

\newcommand{\myscale}{0.7}

\newcommand{\ptp}{{\em peer-to-peer}}

\newcommand{\edk}{{\em eDonkey}}
\newcommand{\udp}{{\sc udp}}
\newcommand{\tcp}{{\sc tcp}}
\newcommand{\ip}{{\sc ip}}

\newcommand{\mdc}{{\sc md5}}
\newcommand{\mdq}{{\sc md4}}
\newcommand{\xml}{{\sc xml}}
\newcommand{\pcap}{{\sc pcap}}

\newcommand{\inet}{internet}

\newcommand{\ie}{{\em i.e.}}
\newcommand{\fid}{{\em fileID}}
\newcommand{\cid}{{\em clientID}}

\newcommand{\eg}{{\em e.g.}}
\newcommand{\edonkey}{{\em eDonkey}}

\newcommand{\nbqueries}{8\,867\,052\,380}
\newcommand{\nbip}{89\,884\,526}
\newcommand{\nbhash}{275\,461\,212}
\newcommand{\PCAPlosses}{250\,266}
\newcommand{\PCAPcount}{31\,555\,295\,781} % meme chose qu'IPcount ?

\newcommand{\UDPcount}{14\,124\,818\,158}

\newcommand{\EDKcount}{8\,867\,052\,380}

\author{Frederic Aidouni, Matthieu Latapy and Clemence Magnien\\
\affaddr{LIP6 -- CNRS and University Pierre \& Marie Curie}\\
\affaddr{104 avenue du president Kennedy, 75016 Paris, France}\\
\email{Firstname.Lastname@lip6.fr}
}
\title{Ten weeks in the life of an eDonkey server
}

\begin{document}
\maketitle
\begin{abstract}
This paper presents a capture of the queries managed by an {\edk} server during almost 10 weeks, leading to the observation of almost 9 billion messages involving almost 90 million users and more than 275 million distinct files. Acquisition and management of such data raises several challenges, which we discuss as well as the solutions we developed. We obtain a very rich dataset, orders of magnitude larger than previously avalaible ones, which we provide for public use. We finally present basic analysis of the obtained data, which already gives evidence of non-trivial features.
\end{abstract}

\footnotetext{This paper is candidate to the best paper award.}

\section{Introduction}

Collecting live data on running {\ptp} networks is an important task to grasp their fundamental properties and design new protocols \cite{1217970,saroiu02measurement,DBLP:conf/teletraffic/SaddiG07,europar2004gauron,iptps05guillaume}. To this end, {\edk} is appealing: it is one of the currently largest and most popular {\ptp} systems. Moreover, as it is based on servers in charge of file and source searches, it is possible to capture the traffic of such a server to observe the queries it manages and the answers it provides.

\medskip\noindent
{\bf Contribution and context.}

We describe here a continuous capture of \udp/\ip\ level traffic on an important \edonkey\ server during almost ten weeks, from which we extract the application-level queries processed by the server and the answers it gave. This leads to the observation of \nbqueries\ queries, involving \nbip\ distinct \ip\ addresses and \nbhash\ distinct \fid. We carefully anonymise and preprocess this data, in order to release it for public use and make it easier to analyse. Its huge size raises unusual and sometimes striking challenges (like for instance counting the number of distinct \fid\ observed), which we address.

The obtained data surpasses previously available ones regarding several key features: its wide time scale, the number of observed users and files, its rigorous measurement, encoding, and description, and/or the fact that it is released for public use. It also has the distinctive feature of dealing with user behaviors, rather than protocols and algorithms, or traffic analysis, \eg\ \cite{acosta07trace,legout06rarest,saddi07measurement,karagiannis04transport,tutschku04measurement}. To this regard, it is more related to previous measurement-based studies of peer behaviors in various systems, \eg\ \cite{hughes06deviant,giovann07availability,zghaibeh07impact,handurukande06peer,lefessant04clustering}, and should lead to more results of this kind.

As a passive measurement on a server, it is complementary of passive traffic measurements in the network \cite{karagiannis04transport,tutschku04measurement,saddi07measurement}, and client-side passive or active measurements \cite{zghaibeh07impact,handurukande06peer,lefessant04clustering} previously conducted on \edonkey. Up to our knowledge, it is the first significant dataset on \edonkey\ exchanges released so far (though \cite{iwdc2004leblond,iptps05guillaume} use similar but much smaller data), and it is the largest \ptp\ trace ever released. Of course, it also has its own limitations (for instance, it does not contain any information on direct exchanges between clients).

\section{Measurement}

Since our goal was to observe {\em real-world} exchanges processed by an \edonkey\ server, we had to capture the traffic on an existing server (with the authorization of its administrator and within legal limits). In this context, it was crucial to avoid any significant overload on neither the server itself nor its administrator. Likewise, installing dedicated material (\eg\ a {\sc dag} card) was impossible.

Moreover, it is of prime importance to ensure a high level of anonymisation of this kind of data. This anonymisation must be done in real-time during the capture. As \ip\ addresses appear at both \udp/\ip\ and \edonkey/appli\-ca\-tion levels, this implies that the network traffic must be decoded to application-level traffic in real-time.

Finally, we want the released data to be as useful for the community as possible, and so we want to format it in a way that makes analysis easier. This plays an important role in our encoding strategy described in Section~\ref{sec-anon}, with a strong impact on data usability which we illustrate in Section~\ref{sec-analysis}.

\begin{figure}[h!]
\centering
\includegraphics[scale=0.8]{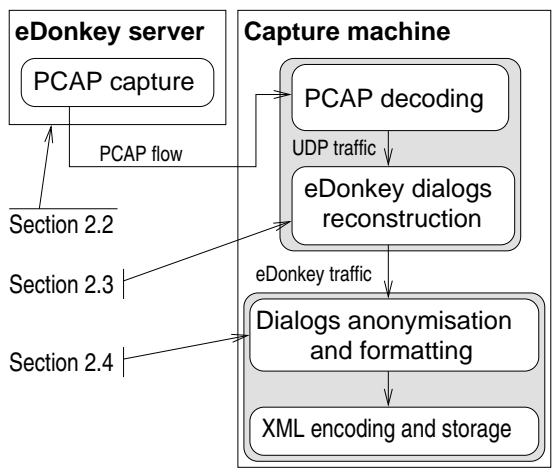}
\caption{{\bf From {\pcap} raw traffic to {\xml} representation}}
\label{fig_data_flow}
\end{figure}

In order to reach these goals, we set up a measurement procedure in three successive steps, as illustrated in Figure~\ref{fig_data_flow}. First, we {\em capture} the network traffic of an \edk\ server using a dedicated program and send it to our capture machine (Section~\ref{sec-capture}). Then this traffic is reconstructed at {\ip} level and {\em decoded} into \edk-level traffic, \ie\ queries and corresponding answers (Section~\ref{sec-ue}). Finally, these queries are {\em anonymised and formated} (Section~\ref{sec-anon}) before being stored as {\xml} documents.

\subsection{The eDonkey protocol briefly}
\label{sec-edk}

{\edk} is a semi-distributed {\ptp} file exchange system based on
directory servers. These servers index files and users, 
and their main role is to answer
to searches for files (based on metadata like filename, size or filetype
for instance), and searches for providers (called {\em sources}) of
given files.

Files are indexed using a {\mdq} hash code, the {\fid}, and are characterised by at least two metadata:
name and size. Sources are identified by a {\cid}, which is their {\ip} address if they are directly reachable
or a 24 bits number otherwise.

{\edk} messages basically fit into four families: management
(for instance queries asking a server for the list of other servers it is aware of);
file searches based on metadata, and the server's answers consisting of a list
of {\fid} with the corresponding names, sizes and other metadata; source searches based on \fid, and the server's answers consisting of a list of sources
(providers) for the corresponding files; and announcements from clients
which give to the server the list of files they provide.

An unofficial documentation of the protocol is available \cite{kulbak05emule}, as well as source code of clients;
we do not give more details here and refer to this document for further information.

\subsection{Traffic capture}
\label{sec-capture}
Before starting any traffic capture, one has to obtain the
agreement of a server administrator. The following guarantees made it possible to reach such an agreement:
negligible impact of the capture on the system;
use of collected data for scientific research; and high level of anonymisation (higher than requested by law).

The ideal solution would be to patch the server source code to add a traffic recording layer.
However, as this source code is {\em not} open-source, this was
impossible. We thus had to design a traffic capture system at the {\ip} level,
then decode this traffic into {\edk} messages.

The server is located in a {\em datacenter} to which we have no access.
A dedicated traffic interception hardware installation was
therefore impossible, and we had to build a software solution. To this end, we
used {\em libpcap}\,\footnote{\url{http://tcpdump.org}}, a standard
{\em ethernet} capture library. We sent a copy of the traffic to a capture
machine, in charge of decoding (Section \ref{sec-ue}), anonymising
(Section \ref{sec-anon}) and storing.

\begin{figure}[h!]
\centering
\includegraphics[scale=\myscale]{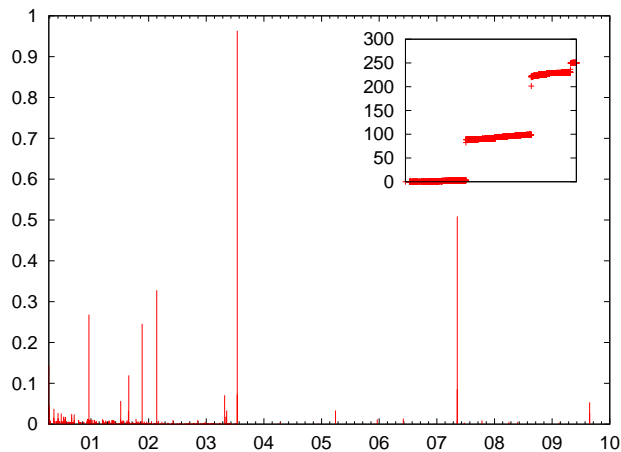}
\caption{
Ethernet packet losses per second during the captureand cumulative losses in thousands of packets (inset). Horizontal axes are labelled by the number of weeks elapsed since the beginning of the measurement. By its end, \PCAPlosses\ packets were lost and \PCAPcount\ were captured.
}
\label{fig_losses}
\end{figure}

This approach leads to packet losses during the capture, due to
the duration of the capture and the network's bandwidth. Indeed, {\em
  libpcap} uses a buffer where the kernel stores captured packets.
In case of traffic peaks, this buffer may be unsufficient and get
full of packets, while some others still arrive. 
The kernel cannot store these new packets in the buffer, and
some are thus lost. The number of lost packets is stored in a kernel
structure, and thus we know the amount of losses that occured, see
Figure \ref{fig_losses}. These losses, although very rare,
make {\tcp} flows reconstruction very
difficult, as packets are missing inside flows\,\footnote{Even without
  packet losses, {\tcp} conversation reconstruction is not an easy
  task, as the server receives about 5000 {\sc syn} packets per
  minute.}. In this paper, we therefore focus on {\udp} traffic only, which
constitutes about half of the captured traffic.

\subsection{From UDP to eDonkey}
\label{sec-ue}

At {\udp} level, our decoding software checks packets and re-assembles the
traffic.
Among \UDPcount\ \udp{} packets captured, 2\,981 are
fragments and 169 are not well-formed.
This corresponds to 949\,873\,704
\edk{} messages, which are then decoded.

The captured
traffic is generated by many poorly reliable clients of different
kinds (and versions), with their own interpretation of the protocol.
Moreover, their source codes are intricate, and the protocol embeds complex
encoding optimisations. Finally, decoding the
server traffic is much harder than programming a client,
and requires an important work of manual decoding of the messages.

Our decoder operates in two steps: a structural validation of messages
(based on their expected length, for example), then, if successful,
an attempt at effective decoding. 
Among the  949\,873\,704 handled \edk{} messages, only 0.68\% were not decoded by our system
(78\% of these messages were structurally incorrect, and thus not decodable).

\subsection{Anonymisation and formating}
\label{sec-anon}

Anonymisation of {\inet} traces is a subtle issue in itself
\cite{allman07issues}.
Since we want to provide the obtained data for public use, we need a very strong anonymisation scheme:
{\cid}, {\fid}, search strings, filenames and filesizes must all
be anonymised (each with a dedicated method, described below). In
addition, timestamps are replaced by the time elapsed since the beginning of the capture
to further limit the
desanonymisation risks.

Filesizes are stored in kilo-bytes (originally they were in bytes);
this precision reduction seems enough to protect this information,
which raises no important privacy issue. 
Search strings, filenames, and server descriptions
are encoded by their {\mdc} hash code, which provides satisfying
anonymisation while keeping a coherent dataset.

Anonymising {\cid} with a hash code is not satisfactory: if one knows the
hash function, it is easy to find the original {\cid} by applying
the function to the $2^{32}$ possible {\cid}. Shuffling strategies are
not strong enough either for this very sensitive data. We therefore
chose to encode {\cid} according to their order of appearance in the captured data: the first one
is anonymised with the value 0, the second with 1 and so on. Although computationaly expensive (see below), this
technique has two advantages: it ensures a very strong anonymisation
level and it makes further use of the dataset much easier, as
anonymised {\cid} are integers between 0 and N-1 (if there are N
distinct {\cid}).

To perform this encoding, we must be able to recognise previously encountered (and anonymised) \cid. We must thus
store throughout the capture the set of \cid\ already seen, with their anonymisation. As
each message contains at least one {\cid}, an overwhelming number of
searches (several billions) must be performed in this set, 
as well as millions of insertions. Classical data structures (like hashtables or trees) are
unsatisfactory in this context: they are too slow and/or too space
consuming. Instead, we used the fact that at most $2^{32}$ dictinct
{\cid} exist: we used an array of $2^{32}$ integers (hence of total size 16 giga-bytes),
and stored the
anonymisation of each \cid\ in the \cid-th cell of this
array. 
This has a high cost in central memory, but allowed us to
anonymise {\cid} with a direct memory access operation only, hence
very efficiently.

\begin{figure}[!h]
\includegraphics[scale=\myscale]{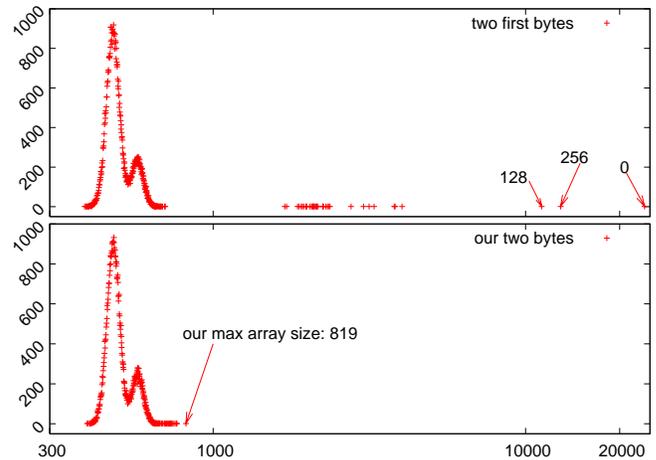}
\caption{{\bf Size distribution of \fid\ anonymisation arrays after one week of capture.} One can observe abnormally large arrays when the arrays are indexed by the first two bytes (array 0 contains 24\,024 elements in this case); using other bytes reduces this significantly.}
\label{fig-dist}
\end{figure}

We also chose to
anonymise the {\fid} by their order of appearance. Here again, the
number of insertions and searches in the corresponding set is huge. As
a consequence, classical set structures were not relevant in this case either. Moreover,
because of the size of {\fid} (128 bits), we could not use the same
solution as for {\cid}.

A possible solution could be to use a sorted array containing {\fid}, with their
anonymisation key. Arrays are compact structures, and when sorted a
dichotomic search is very fast. However, insertion has a prohibitive
cost, due to the reorganisation it implies to keep the array sorted.

One may avoid this problem in a simple way, as {\fid} are hash codes:
they are supposed to be uniformally distributed in their coding space.
As a consequence, dividing the main array in equally-sized smaller ones, indexed by any
part of the \fid, should reduce their size uniformally and thus significantly speed up element insertions.

In our particular situation, dividing the array size by a factor of
65\,536 by using the two first bytes to index 65\,536 arrays seems a
good solution: as we encounter 88 million distinct {\fid} in our
capture, each array length should be around 1500; sorted insertion in
such arrays is reasonable.

However, implementing this strategy led to surprising results: anonymisation arrays
0 and 256 had very large sizes, see Figure~\ref{fig-dist}.
This shows that, in practice, a majority of {\fid} start with
0 or 256, and thus reveals the massive presence of forged {\fid}
\cite{LEE06}. They induce the unbalanced sizes of our anonymisation arrays,
which strongly hampers our computations.

We solved this problem by selecting two different bytes in the
{\fid} to index our 65\,536 arrays. Figure~\ref{fig-dist}
shows that this approach does not perfectly remove the heterogeneity
of array sizes, but it was sufficient for our application.

Finally, the processing method we have described is rather space consuming,
but it is able to decode {\udp} traffic in real-time, which
is crucial in our context.

\subsection{Final dataset}

The final dataset we obtain consists in a series of \EDKcount\ \edk\ messages (queries from clients and answers to these queries from the server) in \xml\ format\,\footnote{We chose \xml\ as output format because it leads to easy-to-read and rigorously specified text files, and, once compressed, does not have a prohibitive space cost.}. It contains very rich information on users at \nbip\ distinct \ip\ addresses dealing with \nbhash\ distinct \fid, while preserving the privacy of users.

This dataset is publicly available with its formal specification\,\footnote{\url{http://www-rp.lip6.fr/~latapy/P2P_data/}}.

\section{Basic analysis}
\label{sec-analysis}

We present in this section a few basic analysis of the data obtained above. Thanks to our formating, the computations needed to obtain these results have a reasonable cost. They give more detailed insight on our dataset. Notice however that these statistics are subject to measurement bias \cite{stutzbach06unbiased}, and only reflect the content of our data; more careful analysis should be conducted to derive accurate conclusions on the underlying objects.

%RF : fid cid : le cid demande le fid (getsource)
%SF : fid cid : ce cid fournit ce fid
%
%RC : cid fid : ce cid demande ce fid
%SC : cid fid : ce fid est fourni par ce cid
%
%so :
% RF -> distrib du nb de clients demandant chaque donnée
% SF -> distrib du nb de clients fournissant chaque donnée
% RC -> distrib du nb de fichiers demandés par chaque pair
% SC -> distrib du nb de fichiers fournis par chaque pair

\subsection{File point of view}

%\begin{figure*}[!ht]
%set term fig color fontsize 12 thickness 1
%set logscale xy
%set output "1.RC.merged.distrib.fig"
%plot [.9:] [.9:] "1.RC.merged.distrib" with points lt 26 notitle
%\centering
%\includegraphics[scale=\myscale]{1.SF.merged.distrib.eps}
%\includegraphics[scale=\myscale]{1.RF.merged.distrib.eps}
%\includegraphics[scale=\myscale]{file.corr.eps}
%\caption{
%{\bf File statistics.} Left: distribution of the number of clients providing each file, \ie\ for each value $x$ on the horizontal axis the number of files provided by $x$ clients. Center: distribution of the number of clients asking for each file. Right: corellations between the number of clients providing and searching for each file, observed by plotting, for a random sample of 100\,000 files, a point at coordinates $x$ and $y$ for each file provided by $x$ clients and asked by $y$.
%}
%\label{fig_files}
%\end{figure*}

\begin{figure}[!ht]
\centering
\includegraphics[scale=\myscale]{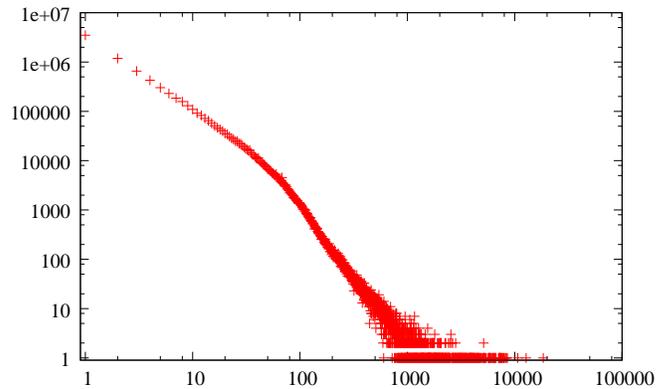}
\caption{
Distribution of the number of clients providing each file, \ie\ for each value $x$ on the horizontal axis the number of files provided by $x$ clients.
}
\label{fig_files1}
\end{figure}

\begin{figure}[!ht]
\centering
\includegraphics[scale=\myscale]{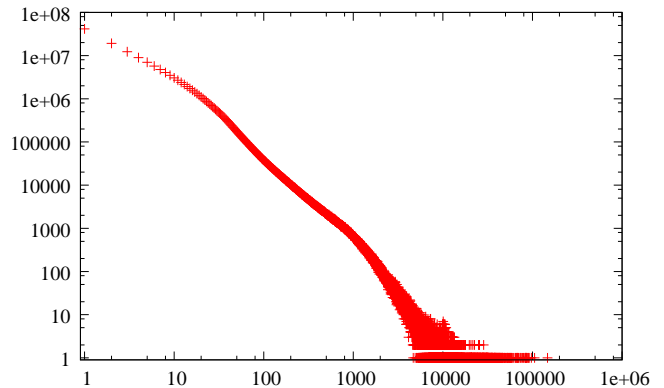}
\caption{
Distribution of the number of clients asking for each file, \ie\ for each value $x$ on the horizontal axis the number of files searched by $x$ clients.
}
\label{fig_files2}
\end{figure}

Figures~\ref{fig_files1} and~\ref{fig_files2} present statistics from the file point of view. They clearly confirm the well known fact that these objects have a very heterogeneous nature: the number of clients providing each file spans several orders of magnitude, as does the number of clients asking for each file. In particular, some files are provided by more than 10\,000 clients, and some are searched by almost 150\,000, which is a non-neglectible fraction of all clients observed\,\footnote{This kind of statistics may be used to conduct audience estimations for the files under concern, most probably audio files or movies.}. On the other hand, a huge amount of files are provided by very few clients (more than 3.5 millions are provided by only one client, and more than one million by two clients only).

The decrease of the distribution of the number of clients providing each file is reasonably well fitted by a power-law (see Figure~\ref{fig_files1}), and the number of clients asking for each file too. This captures the intrinsinc heterogeneity of files regarding the number of clients providing or searching them. This has important consequences on modeling and simulation, as there can be no notion of {\em average} client.

Going further, notice that a better fit would be obtained using a combination of several power-laws, or more subtle laws. This may indicate that files of different nature coexist in the system, which is indeed true (for instance, audio file vs movies, or pornographic content vs classical one). Our data may help in ivestigating this, but this is out of the scope of this paper.

\subsection{Client point of view}

%\begin{figure*}[!ht]
%\centering
%\includegraphics[scale=\myscale]{1.SC.merged.distrib.eps}
%\includegraphics[scale=\myscale]{1.RC.merged.distrib.eps}
%\includegraphics[scale=\myscale]{client-corr.eps}
%\caption{
%{\bf Client statistics.} Left: distribution of the number of files provided by each client, \ie\ for each value $x$ on the horizontal axis the number of clients providing $x$ distinct files. Center: distribution of the number of files each client asks for. Right: corellations between the number of files provided and searched by each client, observed by plotting, for a random sample of 100\,000 clients a point at coordinates $x$ and $y$ for each client providing $x$ files and asking for $y$.
%}
%\label{fig_clients}
%\end{figure*}

\begin{figure}[!ht]
\centering
\includegraphics[scale=\myscale]{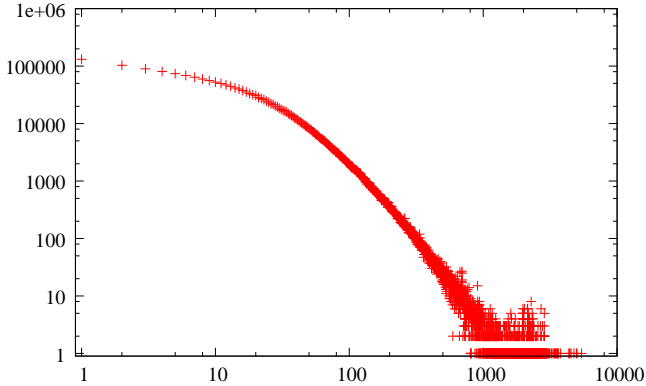}
\caption{
Distribution of the number of files provided by each client, \ie\ for each value $x$ on the horizontal axis the number of clients providing $x$ distinct files.
}
\label{fig_clients1}
\end{figure}

\begin{figure}[!ht]
\centering
\includegraphics[scale=\myscale]{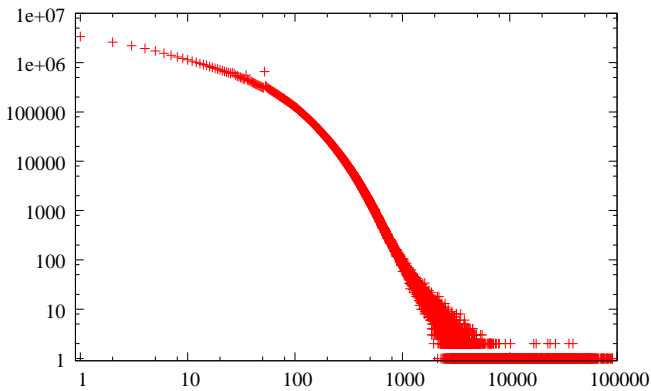}
\caption{
Distribution of the number of files each client asks for, \ie\ for each value $x$ on the horizontal axis the number of clients searching $x$ distinct files.
}
\label{fig_clients2}
\end{figure}

Similarily, Figures~\ref{fig_clients1} and~\ref{fig_clients2} present statistics from the client point of view. They also confirm that clients are very heterogeneous regarding the number of files they provide or search for: both numbers span several orders of magnitudes, with clients providing more than 5\,000 files and/or searching for almost one hundred of thousand files, while hundreds of thousand clients provide or search only a few files. This accounts for the high heterogeneity of user behaviors regarding their use of \ptp\ systems.

Notice however that these distribution are far from power-laws. The number of provided files would not be fitted for small values, and the number of files asked for clearly has several regimes (a slow slope at the beginning, then a sharper one, and a wide range of values with only few occurrences). This may reveal different kinds of activity, and in particular some clients scanning the network to identify many file sources (which is also indicated by the inhomogeneous repartition of \fid\ observed in Section~\ref{sec-anon}). One may investigate this further by observing the correlations between the number of files provided and asked for, for instance, but this is out of the scope of this paper.

Finally, we observe that the distribution of the number of files provided by each client (Figure~\ref{fig_clients1}) indicates an unexpected large number of clients providing a few thousands of files. This may be due to limitations in client software, like for instance a maximal number of files manageable in a same directory on some systems.

Likewise, the distribution of the number of files asked by each client displays a surprisingly singular value: there is a clear peak for the number of peers asking for 52 files. This may be due to a maximal number of queries allowed by a widely used client software.

\subsection{Other statistics}

\begin{figure}[h!]
\centering
\includegraphics[scale=\myscale]{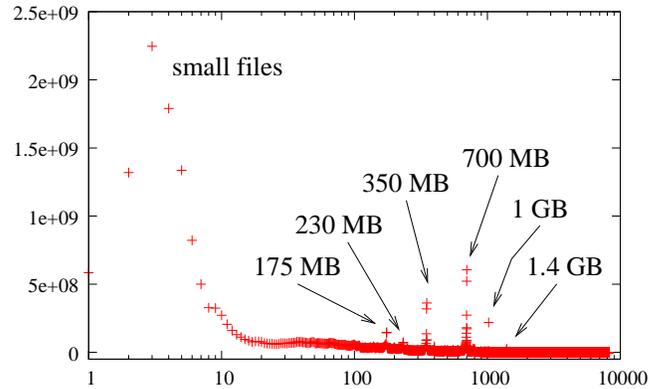}
\caption{File size distribution, \ie\ for each encountered file size (horizontal axis) the number of files having this size (vertical axis).}
\label{fig_size}
\end{figure}

Many other statistics may be observed. For instance, we display in Figure~\ref{fig_size} the distribution of the size of exchanged files (the answers of the server to some queries indicate the size of found files). One observes many small files (probably music files), and clear peaks at 700 MB (typical size of a CD-ROM), and at fractions (1/2, 1/3, 1/4) or multiples (2 $\times$) of this value. The peak at 1 GB may indicate that users split very large files (DVD images for instance) into 1 GB pieces.

This plot reveals the fact that, even though in principle files exchanged in P2P systems may have any size, their actual sizes are strongly related to the space capacity of classical exchange and storage supports.

\section{Conclusion}

This paper presents a capture of the queries managed by a live \edonkey\ server at a scale significantly larger than before, both in terms of duration, number of peers observed, and number of files observed. This dataset is available for public use (with its formal specification) in an easy-to-use and rigorous format which significantly reduces the computational cost of its analysis. We present a few basic analysis which give more information on the collected data.

This work may be extended by conducting measurements of \tcp\ \edonkey\ traffic, and more generally by measuring the \edonkey\ activity using complementary methods (active measurements from clients, for instance). The measurement duration may also be extended even more, and likewise the traffic losses may be reduced.

From an analysis point of view, this work opens many directions for further research. For instance, it makes it possible to study and model user behaviors, communities of interests, how files spread among users, etc. Most of these directions were out of reach with previously available data, and they are crucial from both fundamental and applied points of view.

\bibliographystyle{abbrv}
\bibliography{article,pam}

\end{document}